\documentstyle[multicol,prb,aps,epsf]{revtex}
\begin{document}

\title{
Weak-Coupling Approach to Hole-Doped $S=1$ Haldane Systems
}
\author{Satoshi Fujimoto$^1$ and Norio Kawakami$^2$}
\address{
$^1$Department of Physics, Faculty of Science,
Kyoto University, Kyoto 606, Japan\\
$^2$Yukawa Institute for Theoretical Physics,
Kyoto University, Kyoto 606, Japan
}
\maketitle
\begin{abstract}
As a weak-coupling analogue of hole-doped $S=1$ Haldane
systems, we study two models for coupled chains via Hund
coupling; coupled Hubbard chains,
and a Hubbard chain coupled with an $S=1/2$ Heisenberg chain.
The fixed point properties of these models are investigated by
using bosonization and renormalization group methods.
The effect of randomness on these fixed points is also studied.
It is found that the presence of the disorder parameter inherent
in the Haldane state in the former model
suppresses the Anderson localization for weak
randomness, and stabilizes the Tomonaga-Luttinger liquid state
with the spin gap.
\end{abstract}
\begin{multicols}{2}
The Heisenberg spin chain with integer spin shows a
remarkable feature, so-called the Haldane gap\cite{halg}.
Recently, hole-doping into $S=1$ Haldane gap systems was
realized in ${\rm Y_2BaNiO_5}$\cite{dt}. It inspires
theoretical interest in the effects of carrier-doping
into Haldane systems, which have not been studied well
so far\cite{penc}.
In this paper we consider weak-coupling models for
hole-doped Haldane systems (HDHS); coupled Hubbard chains
via Hund coupling (referred to as model I)
and a Hubbard chain coupled with an
 $S=1/2$ Heisenberg chain (model II). The model I
corresponds to the case where the energy levels of
electrons which compose the $S=1$ state
are almost degenerate, whereas the model II
to the case where these levels are largely separated,
and the lower level can be considered to be localized.
Coupled chain problems have been studied extensively
in several contexts
\cite{drs,grs,kv,sm,fk,sch,hd,wt,ps}.
The model I is somehow related to the models studied in
\cite{drs,grs,kv,sm,sch,hd,wt,ps}, and the model II
is the Kondo lattice model considered in \cite{fk}
with ferromagnetic coupling.
Applying Abelian and non-Abelian bosonization methods,
we investigate
low-energy properties of these models, and find that
they show quite different behaviors at the fixed points.
We also study the effects of randomness
on these fixed points,
which may play a crucial role for HDHS
because the hole-doping inevitably induces randomness
to the system.
By using replica trick and renormalization group analysis,
we find that the quantum disordered state
inherent in Haldane systems suppresses the Anderson
localization for weak randomness.

We first consider the model I without randomness.
The Hamiltonian is given by
\begin{eqnarray}
H&=&-t\sum_{i\sigma} c^{\dagger}_{i\sigma}c_{i+1\sigma}+h.c.
+U\sum c^{\dagger}_{i\uparrow}c_{i\uparrow}
c^{\dagger}_{i\downarrow}c_{i\downarrow}
\nonumber \\
& &-t\sum_{i\sigma} d^{\dagger}_{i\sigma}d_{i+1\sigma}+h.c.
+U\sum d^{\dagger}_{i\uparrow}d_{i\uparrow}
d^{\dagger}_{i\downarrow}d_{i\downarrow}
\nonumber \\
& &+J\sum_i \bbox{S}_{c, i}\cdot\bbox{S}_{d, i},
\label{eqn:ham1}
\end{eqnarray}
where $c_{i,\sigma}$ and $c^{\dagger}_{i,\sigma}$ etc.,
are the annihilation and
creation operators of electrons, $U$ is on-site coulomb
interaction in each chain, and
$\bbox{S}_c=c^{\dagger}_{i,\alpha}
\bbox{\sigma}_{\alpha,\beta}c_{i,\beta}/2$
with the Pauli matrix $\bbox{\sigma}$, etc.
Here the last term is the Hund coupling interaction with $J<0$.
At half-filling, in the strong coupling
limit $J\rightarrow -\infty$,
the model reduces to the $S=1$ Heisenberg chain which possesses
the Haldane gap.
Even for small $\vert J\vert$, the system has a finite gap,
and belongs to the universality class of the
Haldane state \cite{sm,hd,wt}.
We study low-energy properties of this model
away from half-filling by applying
standard Abelian bosonization methods.
We first linearize the dispersion of electrons
and express the electron operators in terms of boson fields
$\phi_{c\sigma}$ and $\theta_{c\sigma}$, etc. which
satisfy the commutation relation
$[\phi_{c\sigma}(x),\theta_{c\sigma}(x^{'})]
=i\Theta(x-x^{'})$\cite{emery,hal}.
Introducing the boson fields,
$\phi_{\pm}=(\phi_c\pm\phi_d)/\sqrt{2}$ for the spin sector
and $\phi^{(c)}_{\pm}=(\phi^{(c)}_c
\pm\phi^{(c)}_d)/\sqrt{2}$ for the charge sector
where $\phi_a=(\phi_{a\uparrow}-\phi_{a\downarrow})/\sqrt{2}$
and $\phi^{(c)}_a=(\phi_{a\uparrow}
+\phi_{a\downarrow})/\sqrt{2}$ with $a=c,d$,
and their canonical conjugate momenta $\Pi_{\pm}$
 and $\Pi_{\pm}^{(c)}$,
we can write down the bosonized Hamiltonian as
\begin{eqnarray}
H&=&H_c+H_s \nonumber \\
H_c&=&\sum_{\nu =+,-}\int dx\biggl
[\frac{v^{(c)}_{\nu}}{2 K^{(c)}_{\nu}}
(\partial_x \phi_{\nu}^{(c)})^2+\frac{v^{(c)}_{\nu}
K^{(c)}_{\nu}}{2}\Pi^{(c)}_{\nu}\biggr] \nonumber \\
H_s&=&\sum_{\nu=+,-}\int dx\biggl[\frac{v_{\nu}}{2 K_{\nu}}
(\partial_x \phi_{\nu})^2+\frac{v_{\nu}K_{\nu}}{2}
\Pi_{\nu}\biggr] \nonumber \\
& &+\frac{\rm const.}{\alpha^2}
\int dx [J_1\cos\sqrt{4\pi}\phi_{-}\cos\sqrt{4\pi}\theta_{-}
         \nonumber \\
& &+J_2\cos\sqrt{4\pi}\phi_{+}\cos\sqrt{4\pi}\theta_{-}]
\nonumber \\
& &+\frac{\rm const.}{\alpha^2}
\int dx (J_3\cos\sqrt{4\pi}\phi_{-}+
2J_4\cos\sqrt{4\pi}\theta_{-} \nonumber \\
& &+J_5\cos\sqrt{4\pi}\phi_{+})\cos\sqrt{4\pi}\phi_{-}^{(c)},
 \label{eqn:hab1}
\end{eqnarray}
where $v_{\nu}^{(c)}$, $v_{\nu}$, $K_{\nu}^{(c)}$ and $K_{\nu}$
are Luttinger liquid parameters
in each sector.
Here we dropped irrelevant terms with oscillating factors and
irrelevant intra-chain backward scattering terms.
We also omit Umklapp
terms because we are concerned with the case away from
half-filling.
The charge mode, $\phi_{+}^{(c)}$, is completely decoupled
and hence it is described by
U(1) Gaussian theory with central charge $c=1$
(Tomonaga-Luttinger liquid)\cite{hal}.
Initially in the renormalization procedure,
 $J_1=J_2=J_3=J_4=J_5=J$.
These coupling constants are renormalized
in different ways. From simple dimensional analysis,
it is easily seen that $J_1$-term is always irrelevant.
In the case of strong correlation limit,
$U\rightarrow \infty$,
the scaling dimension of the field,
$\cos\sqrt{4\pi}\phi_{-}^{(c)}$
is close to $1/2$.
For small $J$, the dimension of $J_5$-term is thus smaller
than 2 and generates spin gap in
$\phi_{+}$ mode and charge gap in $\phi^{(c)}_{-}$ mode.
In this case the mass gap is also open either
in $\phi_{-}$ mode or $\theta_{-}$ mode.
The state with the mass gap in $\phi_{-}$ mode corresponds to
the Ising Neel ordered state
with the non-vanishing order parameter
${\rm lim}_{\vert i-j\vert \rightarrow \infty}
(-1)^{i-j}\langle S^z_iS^z_j\rangle=
\langle\cos\sqrt{\pi}\phi_{+}\rangle
\langle\cos\sqrt{\pi}\phi_{-}\rangle$,
whereas that with the mass gap in $\theta_{-}$ mode to the
quantum disordered state which is a weak-coupling analogue of
the Haldane state characterized by the hidden
string order ${\rm lim}_{\vert i-j\vert \rightarrow \infty}
\langle S^z_i\exp(i\pi\sum_{k=i-1}^{j-1} S^z_k)S^z_j\rangle=
\langle\cos\sqrt{\pi}\phi_{+}\rangle$\cite{dnr,tasa}.
The mechanism for the spin gap formation is similar to
that for the coupled spin chains\cite{sch,hd,wt}.
In the present case, the spin-gap state with
massive $\theta_{-}$ mode
realizes at half-filling  \cite{sm,wt}, and hence
it remains massive away from half-filling, because
$\phi^{(c)}_{-}$ mode is already massive in this
parameter region. Then the fixed point is the metallic
state with the spin gap {\it \'a la} Haldane.
A similar spin-gap state was obtained for {\it t-J} or
Hubbard chains coupled via transverse hopping \cite{drs,grs,kv}.
At this fixed point, dominant correlations
develop in the inter-chain singlet
pairing and ``$4k_{\rm F}$'' CDW.
The order parameters are, respectively, given by
$
O_{SS}(x)=\langle c_{L\sigma}(x)d_{R-\sigma}(x)
+c_{R\sigma}(x)d_{L-\sigma}(x)\rangle,
$
$
O_{CDW}(x)=
\langle c^{\dagger}_{L\sigma}d^{\dagger}_{L\sigma^{'}}
d_{R\sigma^{'}}c_{R\sigma}\rangle.
$
The correlation functions of these order parameters show
algebraic decay;
$\langle O_{SS}(x)O_{SS}(0)\rangle\sim x^{-1/2K^{(c)}_{+}}$
and $\langle O_{CDW}(x)O_{CDW}(0)\rangle
\sim x^{-K^{(c)}_{+}/2}$.
In our case $K^{(c)}_{+}<1$ and hence the fluctuation for
``$4k_{\rm F}$'' CDW is more dominant than that for
singlet superconductivity.

We now discuss the effects of random impurities
on this fixed point.
The effects of randomness on Tomonaga-Luttinger liquid were
investigated by several authors\cite{app,sf,gs}.
Here we apply renormalization group methods used by
Giamarchi and Schulz\cite{gs}.
We introduce the random impurity potential,
\begin{eqnarray}
H_{imp}=\sum_{\sigma}\int dx[\xi(x)(c_{\sigma L}^{\dagger}
c_{\sigma R}+
d^{\dagger}_{\sigma L}d_{\sigma R})+ h.c.],
\end{eqnarray}
with the Gaussian distribution,
\begin{equation}
P_{\xi}=\exp(-D_{\xi}^{-1}\int\xi^{*}(x)\xi(x)dx).
\end{equation}
Here we omit forward scatterings
because they can be incorporated into
the shift of the chemical potential and do not affect
the fixed point properties.
In order to deal with the quenched disorder we use a replica
trick.
We consider the case for weak randomness and incorporate
only the first order
contribution in $D_{\xi}$.
Then there is no coupling between different replica indices,
which will be omitted below.
Applying a standard renormalization group analysis\cite{gs},
 we obtain the scaling equations
up to the lowest order in $J$ and $D_{\xi}$,
\begin{eqnarray}
\frac{{\rm d}D_{\xi}}{{\rm d}l}&=&\biggl(3-\frac{K_{+}}{2}
-\frac{K_{-}}{2}-\frac{K_{+}^{(c)}}{2}
-\frac{K_{-}^{(c)}}{2}\biggr)D_{\xi} \nonumber \\
& &-J_3 D_{\xi}-J_5 D_{\xi},\label{eqn:sca1} \\
\frac{{\rm d}J_3}{{\rm d}l}&=&(2-K_{-}-K^{(c)}_{-})J_3
-D_{\xi}, \label{eqn:sca3} \\
\frac{{\rm d}J_4}{{\rm d}l}&=&(2-\frac{1}{K_{-}}
-K^{(c)}_{-})J_4, \label{eqn:sca4} \\
\frac{{\rm d}J_5}{{\rm d}l}&=&(2-K_{+}-K^{(c)}_{-})J_5
-D_{\xi}, \label{eqn:sca5} \\
\frac{{\rm d}K_{+}}{{\rm d}l}&=&-\frac{v_{+}}{2v_{+}^{(c)}}
D_{\xi}K_{+}^2
-\frac{v_{+}}{2v_{+}^{(c)}}J_5^2K_{+}^2,
\label{eqn:sca6} \\
\frac{{\rm d}K_{-}}{{\rm d}l}&=&-\frac{v_{-}}{2v_{+}^{(c)}}
D_{\xi}K_{-}^2
-\frac{v_{-}}{2v_{+}^{(c)}}J_3^2K_{-}^2
+\frac{2v_{-}}{v_{+}^{(c)}}J_4^2,
\label{eqn:sca7} \\
\frac{{\rm d}K_{-}^{(c)}}{{\rm d}l}&=&
-\frac{v_{-}^{(c)}D_{\xi}K_{-}^{(c)2}}{2v_{+}^{(c)}}
-\frac{v_{-}^{(c)}}{2v_{+}^{(c)}}(J_3^2+4J_4^2+
J_5^2)K_{-}^{(c)2}, \label{eqn:sca9}
\end{eqnarray}
where we did not display the equations irrelevant
to the following discussions, and omitted the renormalization
effects due to $J_2$ term, because they do not change the
intrinsic results.  Since the quantum
disordered spin-gap state realizes in the pure system,
we initially have $2-K_{+}-K_{-}^{(c)}>0$,
$2-K_{-}-K_{-}^{(c)}<0$,
and $2-1/K_{-}-K^{(c)}_{-}>0$.
We can see from eqs.(\ref{eqn:sca6}) and (\ref{eqn:sca9})
that the values of $K_{+}$ and $K_{-}^{(c)}$ are reduced by
the randomness, and hence $2-K_{+}-K_{-}^{(c)}>0$ holds
in the process of the renormalization.
Then from eq.(\ref{eqn:sca5}),
$J_5$ scales to the strong-coupling regime
and gaps open in $\phi_{+}$ and
$\phi^{(c)}_{-}$, resulting in
$K_{+},K_{-}^{(c)}\rightarrow 0$.
Moreover eq.(\ref{eqn:sca3}) indicates the fixed point
value of $J_3$ is $D_{\xi}/(2-K_{-}-K_{-}^{(c)})$, and hence
the terms including $J_3$ in eqs.(\ref{eqn:sca1}) and
(\ref{eqn:sca7}) are negligible to the lowest order
in $D_{\xi}$.
As a result, eqs.(\ref{eqn:sca1}), (\ref{eqn:sca4}),
and (\ref{eqn:sca7}) become
\begin{eqnarray}
\frac{{\rm d}D_{\xi}}{{\rm d}l}&=&\biggl(3-\frac{K_{-}}{2}-
\frac{K_{+}^{(c)}}{2}\biggr)D_{\xi},\label{eqn:msca1} \\
\frac{{\rm d}J_4}{{\rm d}l}&=&\biggl(2-\frac{1}{K_{-}}\biggr)J_4,
\label{eqn:msca4} \\
\frac{{\rm d}K_{-}}{{\rm d}l}&=&-\frac{v_{-}}{2v_{+}^{(c)}}
D_{\xi}K_{-}^2+\frac{v_{-}}{2v_{+}^{(c)}}J_4^2.
\label{eqn:msca7}
\end{eqnarray}
We also note that the correction of $K_{-}$ by randomness
can be neglected in the right-hand side of
eq.(\ref{eqn:msca1}) up to the first order in $D_{\xi}$.
So, $K_{-}$ is solely determined by eqs.(\ref{eqn:msca4})
and (\ref{eqn:msca7}) without the term proportional
to $D_{\xi}$.

\begin{figure}
\centerline{\epsfxsize=7.5cm \epsfbox{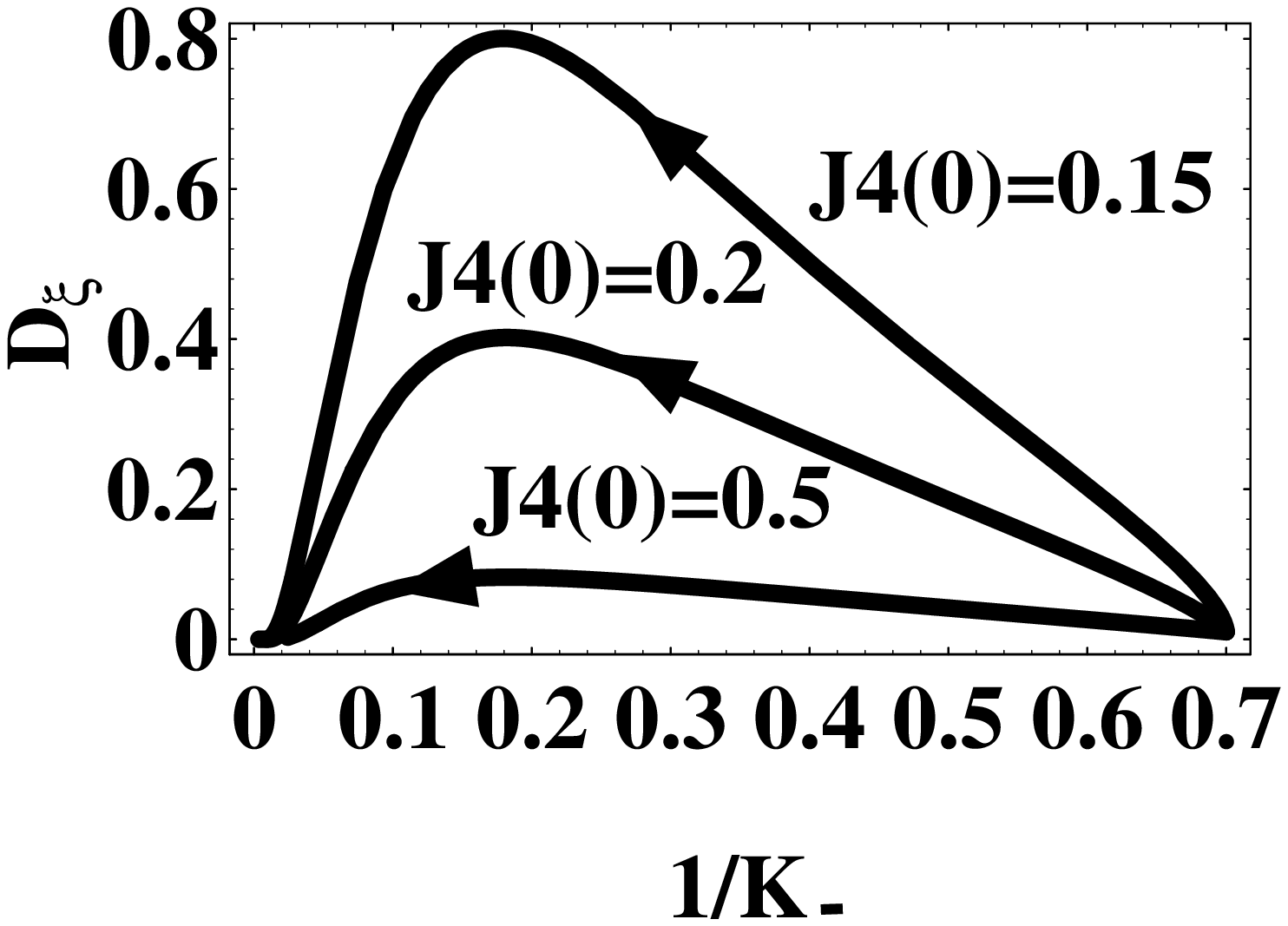}}
{Fig.1.
Plots of $1/K_{-}(l)$ vs $D_{\xi}(l)$ for some initial values
of $J_4$. The initial values of $1/K_{-}$, $D_{\xi}$,
and $K^{(c)}_{+}$ are taken as $0.7$, $0.01$, and $0.6$.}
\end{figure}

\noindent
Solving these equations,
we have $K_{-}\rightarrow \infty$,
because the spin gap in $\theta_{-}$ mode opens
in the absence of randomness.
Thus from eq.(\ref{eqn:msca1}), $D_{\xi}$ scales to $0$ and
the randomness becomes irrelevant.
These arguments are confirmed by the numerical results
for renormalization flows calculated by eqs.(\ref{eqn:msca1})
$\sim$(\ref{eqn:msca7}) (Fig.1).
Note that the suppression of $D_{\xi}$ is due to the spin gap
formation by $J_4$-term, that is,
the presence of the non-vanishing disorder parameter,
$\langle\cos\sqrt{\pi}\theta_{-}\rangle$.
Therefore we can say that the Tomonaga-Luttinger
liquid state in the charge sector is
protected from the transition to the Anderson localization
by the quantum spin disorder.
The physical reason of the suppression of the Anderson
localization may be understood by observing that the presence
of the Haldane gap reduces the number of massless excitations
and hence has a tendency to effectively suppress backward
scatterings due to impurities.

We have checked that for sufficiently large initial values
of $D_{\xi}$ and small
initial values of $J_4$, $D_{\xi}$ scales to a value larger
than unity before $J_4$ scales to the strong-coupling regime,
and our weak-coupling treatment for randomness breaks
down(see also Fig.1).
Although in this case we do not know the fixed point properties,
it seems plausible to expect that sufficiently strong randomness may
destroy the Tomonaga-Luttinger liquid state and bring about the
transition into the Anderson localization state.
It may also leads to the destruction of the quantum disordered
spin-gap state.
Although our discussions here for localization-delocalization
transition rely on weak-coupling renormalization group methods,
we think that the qualitative features should not be changed
even if we take into account higher-order corrections in $J$.


We wish to note that the suppression
of the Anderson localization for weak randomness
is not due to the development of superconducting fluctuation
as found before\cite{gs,sf}, because in the present case
the fluctuation of the ``$4k_{\rm F}$'' CDW is
more dominant than that of superconductivity.
The presence of the quantum spin disorder
inherent in the Haldane state
indeed prevents impurity potential from pinning the CDW.
It may be a new type
of depinning effect of the CDW.
Hence, if a doped-Haldane system belongs to the class of
model I, we can observe the power-law behavior in
correlation functions more easily, owing to the stability of
Tomonaga-Luttinger liquid state against randomness.

We next discuss the model II.
The Hamiltonian is given by
\begin{eqnarray}
H&=&-t\sum_{i\sigma} c^{\dagger}_{i\sigma}c_{i+1\sigma}+h.c.
+U\sum c^{\dagger}_{i\uparrow}c_{i\uparrow}
c^{\dagger}_{i\downarrow}c_{i\downarrow}
\nonumber \\
& &+J_d\sum_i\bbox{S}_{d, i}\cdot\bbox{S}_{d, i+1}
+J \sum_i\bbox{S}_{c, i}\cdot\bbox{S}_{d, i},
\end{eqnarray}
where $J_d$ is the anti-ferromagnetic exchange
interaction ($J_d>0$) and other notations are the same as
those of eq.(\ref{eqn:ham1}).
The charge degree of freedom for $d$-electrons is completely
frozen in this model, which may approximately
describe the system in which the energy
levels of electrons composing the $S=1$ state are
largely separated. In the strong-coupling limit at half-filling,
this Hamiltonian also reduces to the  $S=1$
Heisenberg model. For small $\vert J\vert$, it
can be a weak-coupling analogue
of the Haldane system as in the case of model I.
Note that for $J>0$, this model coincides with the Kondo
lattice model with the nearest-neighbor exchange
interaction between localized spins studied in \cite{fk}

In order to properly describe
symmetry of the system away from half-filling at the fixed point,
non-Abelian bosonization \cite{wit,af1} is
more suitable than Abelian bosonization
because two chains decouple at the fixed point and
SU(2) symmetry of each chain should be retained as we will see below.
By applying non-Abelian bosonization formula to the spin sector,
we can write down the Hamiltonian in a continuum limit
for the case away from half-filling as
\begin{eqnarray}
H&=&H_c+H_s \nonumber \\
H_c&=&\int dx\biggl[\frac{v^{(c)}_c}{2 K^{(c)}_c}
(\partial_x \phi_{c}^{(c)})^2+\frac{v^{(c)}_cK^{(c)}_c}{2}
\Pi^{(c)}_c\biggr] \nonumber \\
H_s&=&\int dx\frac{2\pi v_c}{3}[\bbox{J}_{cL}\cdot\bbox{J}_{cL}+
\bbox{J}_{cR}\cdot\bbox{J}_{cR}] \nonumber \\
& &+\int dx\frac{2\pi v_d}{3}[\bbox{J}_{dL}\cdot\bbox{J}_{dL}+
\bbox{J}_{dR}\cdot\bbox{J}_{dR}] \nonumber \\
& &+J_m\int dx[\bbox{J}_{cL}\cdot\bbox{J}_{dL}+
\bbox{J}_{cR}\cdot\bbox{J}_{dR}] \nonumber \\
& &+J_{ir}\int dx[\bbox{J}_{cL}\cdot\bbox{J}_{dR}+
\bbox{J}_{cR}\cdot\bbox{J}_{dL}], \label{eqn:hab2}
\end{eqnarray}
where $\bbox{J}_{cL}$ and $\bbox{J}_{dL}$, etc. are
the current operators for spinons of $c$- and $d$-electrons,
which satisfy SU(2) Kac-Moody algebra,
and initially we have $J_m=J_{ir}=J$. Here we dropped
irrelevant oscillating terms.
In general, spinons of $c$- and $d$-electrons have
different velocities, $v_c$ and $v_d$. From the operator product
expansion of level-1 SU(2) Kac-Moody
algebra, we obtain the scaling equations for
couplings $J_m$ and $J_{ir}$,
\begin{equation}
\frac{{\rm d} J_m}{{\rm d} l}=0, \qquad\quad
\frac{{\rm d} J_{ir}}{{\rm d} l}=\frac{J^2_{ir}}{2\pi (v_c+v_d)}.
\end{equation}
Thus the last term of $H_s$ is marginally irrelevant
for $J<0$. The $J_m$-term just renormalizes
the velocities of spinons and does not
break SU(2) symmetry in each chain.
Hence at the low-energy fixed point there exist one massless mode in
the charge sector described by $c=1$ Gaussian theory and
two massless modes in the spin sector described by level-1 SU(2)
Wess-Zumino-Witten theory\cite{kz}, and consequently
there is no spin gap away from half-filling.
Therefore doping holes in this model drastically changes
characteristic properties of spin excitations,
in contrast to the model I.
We note that the existence of two massless spin modes
is also contrasted to the Kondo lattice model
with the antiferromagnetic coupling $J>0$ where
only one massless spin mode with a large Fermi surface exists
\cite{ntu,fk}.

At this fixed point, two kinds of
spinons have different pseudo-Fermi surfaces.
Therefore correlation functions involving spin excitations
show singularities at two different points
reflecting the existence of two pseudo-Fermi surfaces.
For example, the spin correlation function for the total spins,
$\bbox{S}(x)=\bbox{S}_c(x)+\bbox{S}_d(x)$, is given by
\begin{equation}
\langle S^{z}(x)S^{z}(0)\rangle \sim
\frac{A_0}{x^2}+\frac{A_1 e^{i2k_{\rm F} x}}{x^{\alpha}}+
\frac{A_2 e^{i\pi x}}{x}+c.c.
\end{equation}
where $A_0, A_1$, and $A_2$ are some constants, and the
exponent of the second term is given by $\alpha=1+K_c^{(c)}$.
Here we omitted logarithmic corrections due to marginal operators.
Thus the form factor of the spin correlation
function may exhibit two structures in the momentum space at
$q=2k_{\rm F}$ and $\pi$.

The effects of randomness on the model II can also be investigated.
In this model, two chains are decoupled at the fixed point.
Then the problem is simply divided into two parts;
the 1D Hubbard model and the 1D $S=1/2$ Heisenberg model
with site randomness. For the former problem, according to
Giamarchi and Schulz\cite{gs},
the Anderson localization takes place for repulsive interaction.
For the latter problem, if the on-site correlation between
localized $d$-electrons is sufficiently
strong, the site randomness does not affect the exchange
interaction $J_d$ so much, and hence the massless spin mode can
survive, although the system  may be of the glass state
if we additionally include randomness for exchange interaction.

Comparing the results for the models I and II, which exhibit
quite different behaviors upon
doping, we can say that  characteristic properties of hole-doped
Haldane systems strongly depend on
whether the energy levels of electrons consisting the $S=1$
state are nearly degenerate, or largely separated.

According to the experimental results of ref.\cite{dt},
low-energy excited states below the spin
gap appear upon doping, and localization occurs
in transport properties. These seem to be consistent with
the model II with random potential.
We think that the model II may effectively describe
low-energy properties of the doped Haldane system
${\rm Y}_{2-x}{\rm Ca}_x{\rm BaNiO}_5$ for larger
doping regions, though hole-doping has been realized in  a way
different from the present model\cite{dt,penc}.

This work was partly supported by a Grant-in-Aid from the Ministry
of Education, Science and Culture.
\end{multicols}
\begin{multicols}{2}

\end{multicols}

\end{document}